\documentclass{aastex631}

\newcommand{\prot}{\ifmmode {P_{\mathrm{rot}}}\else$P_{\mathrm{rot}}$\fi}

\received{June 8, 2022}
\accepted{\today}

\shorttitle{Rotation in Praesepe Tidal Tails}
\shortauthors{McDivitt et al.}
\graphicspath{{./}{figures/}}

\begin{document}

\title{Using stellar rotation to identify tidally stripped members of the Praesepe open cluster
}


\newcommand{\lafayette}{Department of Physics, Lafayette College, 730 High St, Easton, PA 18042, USA}

\newcommand{\columbia}{Department of Astronomy, Columbia University, 550 West 120th Street, New York, NY 10027, USA}

\newcommand{\amnh}{Department of Astrophysics, American Museum of Natural History, Central Park West at 79th Street, New York, NY 10034, USA}

\newcommand{\cunygc}{Physics, The Graduate Center, City University of New York, New York, NY 10016, USA}

\newcommand{\hunter}{Department of Physics and Astronomy, Hunter College, City University of New York, 695 Park Avenue, New York, NY 10065, USA}

\correspondingauthor{Stephanie T.\ Douglas}
\email{StephanieTDouglas@gmail.com}

\author[0000-0001-8356-4642]{Jessica McDivitt}
\affiliation{\lafayette}

\author[0000-0001-7371-2832]{Stephanie T.\ Douglas}
\affiliation{\lafayette}

\author[0000-0002-2792-134X]{Jason Lee Curtis}
\affiliation{\columbia}

\author[0000-0001-9482-7794]{Mark Popinchalk}
\affiliation{\amnh}
\affiliation{\cunygc}
\affiliation{\hunter}

\author[0000-0002-8047-1982]{Alejandro N\'u{\~n}ez}
\altaffiliation{NSF MPS-Ascend Postdoctoral Research Fellow}
\affil{\columbia}

\begin{abstract}

As an open cluster orbits the Milky Way, gravitational fields distort it, stripping stars from the core and forming tidal tails. 
Recent work has identified tidal tails of the Praesepe cluster; we explore rotation periods as a way to confirm these candidate members.
In open clusters, the rotation period distribution evolves over time due to magnetic braking. 
Since tidally stripped stars originally formed within the cluster, they should follow the same period distribution as in the cluster core.  
We analyze 96 candidate members observed by NASA's Transiting Exoplanet Survey Satellite (TESS) mission. We measure reliable rotation periods for 32 stars, while 64 light curves are noise-dominated. 
The 32 newly identified rotators are consistent with the period distribution in the core, and with past membership in Praesepe.
We therefore suggest that for nearby open clusters, stellar rotation offers a quick and inexpensive method for confirming past members dispersed into tidal tails.
\end{abstract}


\section{Introduction}

The precise astrometry from the Gaia mission has enabled the discovery of extended stellar structures in the Milky Way, including both new structures and tidal features associated with known star clusters \citep[e.g.,][]{kounkel2019,meingast2019,roser2019a}. 
Using Gaia DR2, \citet{roser2019} identify 389 stars that may have been tidally stripped from the benchmark Praesepe open cluster. 
Astrometric data alone, however, cannot confirm whether these tidal tails contain true siblings of the remaining cluster core stars.

Because the period distributions of open clusters evolve with age, we can use stellar rotation periods (\prot) to support or rule out candidate tidal tail members. 
Stellar rotation has been successfully employed to confirm the existence of and age-date a variety of young stellar structures including stellar streams \citep[e.g., Pisces--Eridanus, Theia~456:][]{Curtis2019_PscEri, Andrews2022_T456}
and diffuse coronae surrounding open clusters \citep[e.g., NGC~2516:][]{Bouma2021_2516}.
Recently-formed star clusters show a roughly uniform distribution of \prot\  between 1--10~days for stars with masses $0.1 M_\odot \lesssim M_* \lesssim1.3 M_\odot$ \citep[e.g., NGC 6530 and the Orion Nebula Cluster;][]{bouvier2014}. 
This initial distribution evolves so that at a given age, the higher-mass stars in this range tend to follow a single-valued mass--period relationship, while lower-mass stars still have a range of \prot\ values. 
The overall distribution also evolves to slower periods over time. 
Therefore, if candidate tidal tail members follow the same period distribution as in the cluster core, we can assume that they are the same age as and formed along with the core stars. 

To measure rotation periods for possible tidally stripped members of Praesepe, we turn to NASA's Transiting Exoplanet Survey Satellite (TESS) mission. 
We select 96 candidate tidal tail members from \citet{roser2019}, observed in TESS sectors 7, 20, 21, and 34. Many of our targets were observed in two sectors.
We use a Python GUI\footnote{The GUI is also used for the visual inspection, and the associated code is under development at \\ \url{https://github.com/SPOT-FFI/tess_check}.} to extract and detrend light curves from the full-frame images using two methods: Causal Pixel Modeling (CPM) and Simple Aperture Photometry (SAP).  
We use the \texttt{unpopular} package \citep{hattori2021} to produce the CPM light curves, and \texttt{photutils} \citep{bradley2021} to produce the SAP light curves. 

We measure rotation periods using a Lomb--Scargle periodogram before visually inspecting each light curve.
One author inspected all 96 targets twice, selecting the best sector(s) as appropriate and assigning a quality score each time. Ambiguous results were then resolved through discussion with a co-author. 
This left us with 64 light curves that were too noisy or systematics-dominated to measure a period, or which appeared flat with no variability. 
Thirty-two stars show clear periodic variability, and can be assessed for cluster membership. 

\begin{figure}[t!]
\plotone{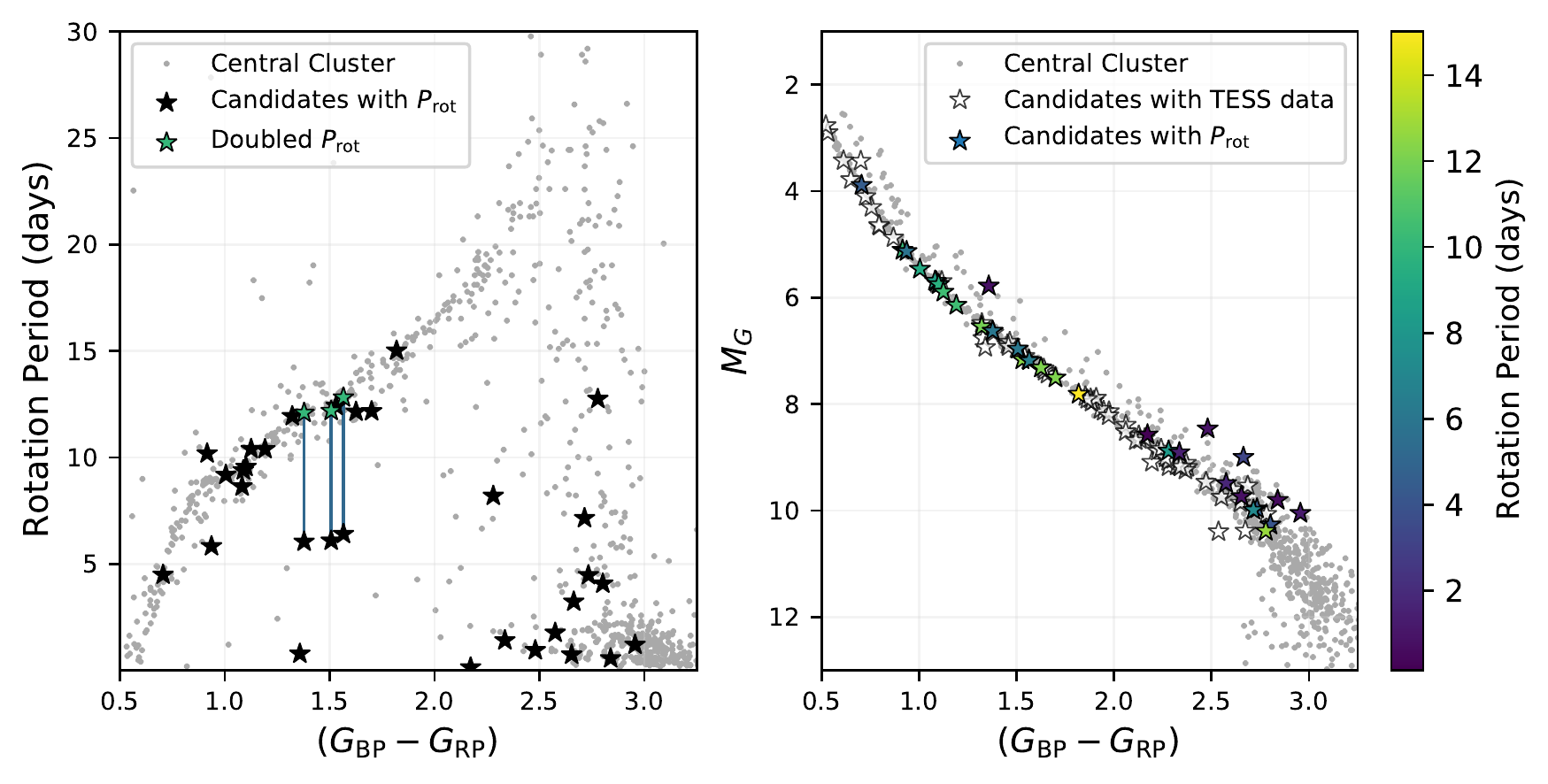}
\caption{(\textit{Left}) Color--period diagram for Praesepe showing that the majority of our newly identified rotators (black stars) are consistent with the \prot\  distribution of the cluster core (gray points). We can conclude that these stars were probably once members of the Praesepe core. Four stars are outliers; three of these become consistent with the cluster core if we double their periods (blue stars). (\textit{Right}) Color--magnitude diagram for Praesepe, showing the locations of candidate tidal tail members with (colored stars) and without (open stars) measured \prot. Core cluster members with measured rotation periods are also shown (grey points). 
 \label{figfig}}
\end{figure}

Figure~\ref{figfig} shows a color--period diagram comparing our 32 new \prot\ to existing \prot\ for cluster core stars. 
The primary sources of these core \prot\ are K2 data analyzed by \citet{douglas2017} and \citet{rampalli2022}. 
We use the same core membership catalog as \citet{rampalli2022}, which is restricted to stars with membership probabilities $P_{\rm mem}\ge70\%$.
Our new TESS \prot, along with Gaia colors and object identifiers, are provided as ``data behind the figure.'' 
Our ability to measure longer periods with TESS is limited by the relatively short sector duration, midway data gaps, and the calibration process that seems to suppress longer signals, 
but overall when we measure a \prot\  for a star, it lines up very well with the core distribution. 
Twenty-eight stars clearly follow the slow-rotator sequence (for $(G_{\rm BP} - G_{\rm RP})\lesssim2$) or the broad distribution of periods for M dwarfs ($G_{\rm BP} - G_{\rm RP}\gtrsim2$).
All stars shown are also consistent with the cluster main sequence in a Gaia $M_G$ vs.\ $(G_{\rm BP} - G_{\rm RP})$ diagram.
Therefore, we find strong evidence that 28 of our 32 new rotators are truly former members of Praesepe. 

There are four notable outliers among our TESS targets: three targets with $(G_{\rm BP} - G_{\rm RP})\approx1.5$ and $\prot\approx6$~d, and a very rapid rotator with $(G_{\rm BP} - G_{\rm RP})\approx$1.4. 
For the three outliers with $\prot\approx6$~d, the TESS sector length and mid-sector data downlink prevent us from confidently identifying a \prot\ value. 
Their light curves are also consistent with $\prot\approx12$~d and a double-dip feature --- in this case, the measured $\prot\approx6$~d would represent a $1/2$ harmonic of the true period. The mid-sector data downlink perfectly interrupts the signal every 12~d, meaning we cannot say for certain whether the true \prot\ for these stars is 12~d. 
If our detections represent a half-period harmonic, however, then these stars would be consistent with the slow rotator sequence in the cluster core. 

Finally, the star with $(G_{\rm BP} - G_{\rm RP}) \approx$1.4 and \prot$\approx$1~d (Gaia DR2/EDR3 5753188583680600320) is likely a binary star.
We measure a \prot\  that would be unusually fast for a single star with the same color, 
which would seem to rule it out as a tidally stripped member. 
It has a Gaia EDR3 RUWE value of 1.386, however, suggesting that it is an unresolved binary system. 
The star is also overluminous in the Gaia $M_G$ vs.\ $(G_{\rm BP} - G_{\rm RP})$ diagram. 
The very rapid rotation suggests it is either a short-period, equal-mass, tidally synchronized binary system, or a hierarchical triple system; in either case, this makes it an interesting target for radial velocity (RV) follow-up. 
We can neither confirm nor rule out this star as a former Praesepe member without RVs, so for now we simply say that it could be consistent with prior membership in the cluster. 

Our study suggests that \citet{roser2019} were successful in identifying real associates of the Praesepe cluster, and demonstrates the utility of \prot\ as supporting information for identifying former cluster members. 
The stars for which we could measure \prot\  are all consistent with the cluster; while \prot\  would let us identify some older field stars, we do not find any, indicating that the underlying catalog is reasonably accurate. 

To conclude, we present 32 new rotation periods for candidate tidal tail members of the Praesepe open cluster. 
Twenty-eight of these stars have \prot\  completely consistent with the cluster core, and we consider them truly former members. 
Four other stars are technically outliers: three of them are consistent with the cluster sequence if we double their measured periods, and one is likely a tidally synchronized binary. 
We cannot rule out these stars as former cluster members at this time, but they could be confirmed with additional data. 
Based on how many of our targets with \prot\  are consistent with the Praesepe cluster core, this supports the idea that these stars are part of the same structure as Praesepe, likely tidal tails (or an extended filament of star formation).

\begin{acknowledgements}

The code used to produce Figure~\ref{figfig} and the ``data-behind-the-figure'' may be found at \url{https://github.com/jessmcdivitt/Praesepe-Tails}.

J.M.\ acknowledges support from the Lafayette College EXCEL Scholars Program. 
S.T.D.\ acknowledges support from the National Aeronautics and Space Administration (NASA) under Grant No.\ 80NSSC19K0377 issued through the TESS Guest Investigator program.
 J.L.C.\ is supported by NSF AST-2009840 and NASA TESS GI grant No.\ 80NSSC22K0299 (G04217).

\facility{TESS, Gaia}

\end{acknowledgements}


\bibliography{references}{}
\bibliographystyle{aasjournal}



\end{document}